\documentclass[journal]{IEEEtran}

\usepackage{hyperref}

\usepackage{textgreek}
\usepackage{cite}

\usepackage{graphicx}

\usepackage{amsmath}
\DeclareMathOperator*{\argmin}{argmin}

\begin{document}

\title{Calibration-Free Relaxation-Based\\Multi-Color Magnetic Particle Imaging}

\author{Yavuz~Muslu*, Mustafa~Utkur, Omer~Burak~Demirel and~Emine~Ulku~Saritas
\thanks{\textbf{Manuscript submitted to IEEE Trans Med Imaging.}}
\thanks{Asterisk indicates corresponding author.}
\thanks{*Y. Muslu is with the Department of Electrical and Electronics Engineering, Bilkent University and National Magnetic Resonance Research Center (UMRAM), Ankara, 06800, Turkey. (e-mail: \href{mailto:ymuslu@ee.bilkent.edu.tr}{ymuslu@ee.bilkent.edu.tr})}
\thanks{M. Utkur and O. B. Demirel are with the Department of Electrical and Electronics Engineering, Bilkent University and National Magnetic Resonance Research Center (UMRAM), Ankara, 06800, Turkey.}
\thanks{E. U. Saritas is with the Department of Electrical and Electronics Engineering, Bilkent University, National Magnetic Resonance Research Center (UMRAM), and Neuroscience Program, Sabuncu Brain Research Center, Bilkent University, Ankara, 06800, Turkey.}}

\maketitle

\begin{abstract}
Magnetic Particle Imaging (MPI) is a novel imaging modality with important applications such as angiography, stem cell tracking, and cancer imaging. Recently, there have been efforts to increase the functionality of MPI via multi-color imaging methods that can distinguish the responses of different nanoparticles, or nanoparticles in different environmental conditions. The proposed techniques typically rely on extensive calibrations that capture the differences in the harmonic responses of the nanoparticles.
In this work, we propose a method to directly estimate the relaxation time constant of the nanoparticles from the MPI signal, which is then used to generate a multi-color relaxation map. The technique is based on the underlying mirror symmetry of the adiabatic MPI signal when the same region is scanned back and forth.
We validate the proposed method via extensive simulations, and via experiments on our in-house Magnetic Particle Spectrometer (MPS) setup at 550 Hz and our in-house MPI scanner at 9.7 kHz. Our results show that nanoparticles can be successfully distinguished with the proposed technique, without any calibration or prior knowledge about the nanoparticles.  
\end{abstract}

\begin{IEEEkeywords}
Magnetic particle imaging, multi-color MPI, nanoparticle relaxation, direct estimation, mirror symmetry
\end{IEEEkeywords}

\IEEEpeerreviewmaketitle

\section{Introduction}

\IEEEPARstart{M}{agnetic} Particle Imaging (MPI) is a new and rapidly developing imaging modality, that shows great potential in terms of resolution, contrast, and sensitivity \cite{Gleich2005, Weizenecker2007, Goodwill2012a, Saritas2013a, Bauer2015}. This imaging technique exploits the nonlinear magnetization response of superparamagnetic iron oxide (SPIO) nanoparticles 
which provides MPI with high-contrast imaging capabilities, without any signal from the human tissue \cite{Gleich2005, KuanLu2013, Zheng2016}. MPI uses combination of three different magnetic fields to obtain an image: a static selection field with a strong gradient is used to create a field-free-point (FFP). Then a sinusoidal drive field is applied to move the FFP and scan the field-of-view (FOV). However, human safety limits 
restrict the size of the FOV covered by the drive field \cite{Schmale2013, Saritas2013b, Schmale2015, Saritas2015, Demirel2017}. Hence, low-frequency focus fields are used to shift the FFP to regions that cannot be covered with the small FOV due to drive field alone \cite{Goodwill2012b, Goodwill2012c, Goodwill2013}.

There are two main image reconstruction methods in MPI. System Function Reconstruction (SFR) requires calibration of the nanoparticle response and the imaging system, which is achieved by a time-consuming calibration scan that records the frequency response of the signal induced by the SPIOs located at all voxel locations in the FOV \cite{Rahmer2009, Knopp2010a, Knopp2010b, Rahmer2012}. The alternative image reconstruction technique, x-space reconstruction, does not require any calibration as it reconstructs the image by gridding the acquired signal to the instantaneous position of the FFP \cite{Goodwill2010, Goodwill2011}. The resulting images, in return, show a version of the SPIO distribution blurred by the point spread function (PSF) of the imaging system. Both of these image reconstruction techniques, with their own advantages and disadvantages, map the spatial distribution of the SPIOs. While MPI's quantitative imaging of SPIO distributions can be an important tool by itself for applications such as cardiovascular imaging and stem cell tracking, separation of the signals acquired from different nanoparticles would increase the functionality of MPI even further.

Rahmer \textit{et al.} have recently demonstrated that different SPIO types can be distinguished via a multi-color MPI technique \cite{Rahmer2015}, and numerous applications have already been shown to benefit from this novel approach. Recently, there has been significant progress in using multi-color MPI for catheter tracking during cardiovascular interventions \cite{Haegele2016}. In such applications, one SPIO type is injected into the blood stream for vessel visualization, while the catheter is coated with a different type of SPIO for tracking purposes. A further advancement of this technique involved simultaneous tracking and steering of the catheter tip via using the the magnetic fields in MPI \cite{Nothnagel2016, Rahmer2017}. More recently, the temperature mapping capability of multi-color MPI has also been demonstrated \cite{Stehning2017}. 

Binding state of the SPIOs has been spectroscopically shown to change relaxation behavior as well \cite{Rauwerdink2010}. In this regard, relaxation mapping can be used to probe cell or protein binding of SPIOs. Moreover, there have been efforts on drug delivery via nanocarriers \cite{Bhaskar2010}, where multi-color MPI can offer additional tracking possibilities. Likewise, multi-color MPI can also be used to identify the characteristics of an environment such as viscosity and temperature, which are shown to affect nanoparticle relaxation behavior significantly \cite{Rauwerdink2009, Weaver2009, Utkur2017}. These conditions can be an important tool for probing the changes in tissue environments, e.g., in the case of hyperthermia treatments. 
All of the aforementioned applications create an immense room for further experimental work and research.
 
To date, multi-color MPI has been realized via both SFR and x-space approaches. In the SFR approach, SPIOs were differentiated based on the differences in their harmonic responses, which were obtained using an extensive calibration procedure performed separately for each SPIO type \cite{Rahmer2015}. For the x-space approach, Hensley \textit{et al.} demonstrated that x-space reconstruction is capable of multi-color reconstruction via differentiating relaxation behaviors of different SPIO types, if multiple measurements at different drive field amplitudes are utilized \cite{Hensley2015}.

In this work, we present a novel, calibration-free multi-color MPI technique for x-space MPI. This technique can generate a relaxation time constant map of SPIOs from a single scan at a single drive field amplitude. The proposed technique takes advantage of the back and forth scanning of a FOV to estimate the relaxation time directly from the MPI signal, without any a priori knowledge about the SPIOs. Here, we demonstrate with extensive simulation results, MPS experiments, and imaging experiments that the proposed method successfully distinguishes multiple SPIO types, without requiring any calibrations.

\section{Theory}

\subsection{MPI Signal}
\begin{figure}[htbp!]
\centering
\includegraphics[width=\linewidth]{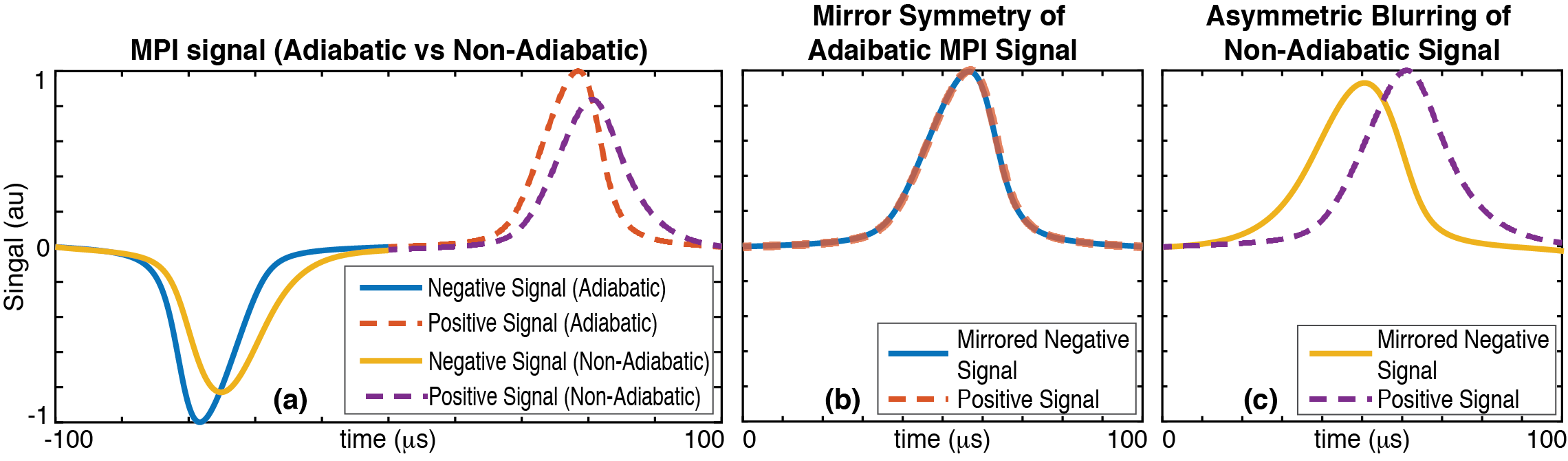}
\caption{The mirror symmetry of the adiabatic MPI signal and how the relaxation effects break that mirror symmetry.
(a) Theoretical MPI signal for a ramp-shaped nanoparticle density. Here, one full period of the MPI signal is shown (i.e., the signals from both the negative and positive cycles are shown). (b) In the adiabatic case, due to the repetitive nature of applied sinusoidal drive field, positive and mirrored negative signals match perfectly. We refer to this phenomenon as "mirror symmetry". (c) The relaxation effect causes an asymmetric blurring of the negative and positive signals, which in turn breaks the mirror symmetry.}
\label{Asymmetric Blurring}
\end{figure}

In x-space MPI, the ideal signal (also known as the adiabatic signal) is defined via the Langevin response of the nanoparticles to an applied drive field \cite{Goodwill2010}:
\begin{equation}
\label{ideal_signal_equation}
s_{adiab}(t)=B_1m\rho(x) * \dot{\mathcal{L}}[kGx]|_{x=x_s(t)} kG\dot{x}_s(t)
\end{equation}
Here, $B_1~[T/A]$ is receive coil sensitivity, $m~[Am^{2}]$ is the magnetic moment of nanoparticles, $\rho(x)~[particles/m^{3}]$ is nanoparticle density, $k~[m/A]$ is nanoparticle property, and $G~[T/m/\mu_{0}]$ is the gradient strength of the selection field. In practice, the MPI signal lags and gets wider due to nanoparticle relaxation effects, i.e., the delay of SPIOs in aligning with oscillating magnetic fields (see Fig. \ref{Asymmetric Blurring}a). In x-space MPI, the relaxation effect is modelled as a temporal convolution of the ideal MPI signal with an exponential kernel \cite{Croft2012}: 
\begin{gather}
s(t)=s_{adiab}(t) * r(t) \label{nonadiabatic_signal_equation} \\
r(t)=\frac{1}{\tau}e^{-\frac{t}{\tau}}u(t) \label{relaxation_kernel}
\end{gather}
Here, r(t) denotes the relaxation kernel and $\tau$ is the relaxation time constant. The resulting MPI signal is called the non-adiabatic signal. Via extensive experimental studies, this simple yet powerful model has been shown to accurately match the relaxation effect for a wide range of drive field frequencies and amplitudes \cite{Croft2012, Croft2015}. 

\subsection{Proposed Method: Direct Relaxation Time Constant Estimation}
In MPI, a sinusoidal drive field is applied to excite the SPIOs, while additional focus fields are applied to control the global positioning of the FFP. For a fixed focus field amplitude, the resulting FFP trajectory scans a partial field-of-view (pFOV) back and forth around a central position. For simplicity, the forward motion of the FFP is called the positive scan, while the backward motion is called the negative scan. For that trajectory, one would expect the signals acquired during these two scans to be mirror symmetric, as displayed in Fig. \ref{Asymmetric Blurring}a-b. However, the nanoparticle relaxation causes an asymmetric blurring that breaks the mirror symmetry (see Fig. \ref{Asymmetric Blurring}c).

We have previously introduced a technique to directly estimate the relaxation time constant from the MPI signal, using the underlying mirror symmetry of the adiabatic signal \cite{Onuker2015, Muslu2016, Utkur2017}. In this technique, we formulated the effects of relaxation in Fourier domain to directly estimate the relaxation time constant. Here, we define $s_{pos}(t)$ as the signal acquired during positive scan, and $s_{neg}(t)$ as the signal acquired during negative scan, both centered with respect to time. For the adiabatic MPI signal, $s_{pos, adiab}(t)$ and $s_{neg, adiab}(t)$ are mirror symmetric, i.e., 
\begin{equation}
\label{adiabatic_pos_neg}
s_{pos, adiab}(t)=-s_{neg, adiab}(-t)=s_{half}(t)
\end{equation}
where, $s_{half}(t)$ denotes half a period of the adiabatic MPI signal. Using the relaxation formulation in Eq. \ref{nonadiabatic_signal_equation}, the non-adiabatic positive and negative signals can be expressed as:
\begin{gather}
s_{pos}(t)=s_{pos, adiab}(t) * r(t)=s_{half}(t) * r(t) \label{nonadiabatic_pos} \\
s_{neg}(t)=s_{neg, adiab}(t) * r(t)=-s_{half}(-t) * r(t) \label{nonadiabatic_neg}
\end{gather}
Here, $s_{half}(t)$ and r(t) are the unknowns of the equations while $s_{pos}(t)$ and $s_{neg}(t)$ are measured waveforms. 
In Fourier domain, these equations can be expressed as:
\begin{gather}
\mathcal{F} \{r(t)\}=R(f)=\frac{1}{1+i2{\pi}f\tau} \label{fourier_kernel} \\
\mathcal{F} \{s_{pos}(t)\}=S_{pos}(f)=S_{half}(f) .R(f) \label{fourier_pos} \\
\mathcal{F} \{s_{neg}(t)\}=S_{neg}(f)=-S_{half}^*(f) .R(f) \label{fourier_neg}
\end{gather}
where $\mathcal{F}$ is the Fourier transform operator, and the time reversal and conjugate symmetry properties of Fourier Transform are used to express $S_{neg}(f)$. Using Eqs. \ref{fourier_kernel}-\ref{fourier_neg} , $\tau$ can be calculated directly as follows:
\begin{equation}
\label{tau_estimtaion}
\hat{\tau}(f)=\frac{S_{pos}^*(f)+S_{neg}(f)}{i2{\pi}f(S_{pos}^*(f)-S_{neg}(f))}
\end{equation}

Ideally, performing this calculation at a single frequency in Fourier domain suffices. However, the accuracy of the estimation is different at each frequency component due to relative strengths of signal vs. noise. To increase the robustness of the estimation against noise, a weighted average of ${\tau}(f)$ is calculated with respect to the magnitude spectrum:
\begin{equation}
\label{tau_estimtaion2}
\hat{\tau}=\frac{\int_{0}^{f_{max}} |S_{pos}(f)|{\tau}(f) \ df }{\int_{0}^{f_{max}} |S_{pos}(f)| \ df}
\end{equation}
Here, $f_{max}$ is an upper threshold for the range of frequencies used, such that $0<f_{max}\ll\frac{F_s}{2}$, where $F_s$ is the sampling frequency. Typically, including frequencies up to $6^{th}$ or $7^{th}$ harmonic of the fundamental frequency suffices, as the signal falls of rapidly with increasing frequency. 

\subsection{Verification with 1D MPI Simulations}
\begin{figure}[htbp!]
\centering
\includegraphics[width=\linewidth]{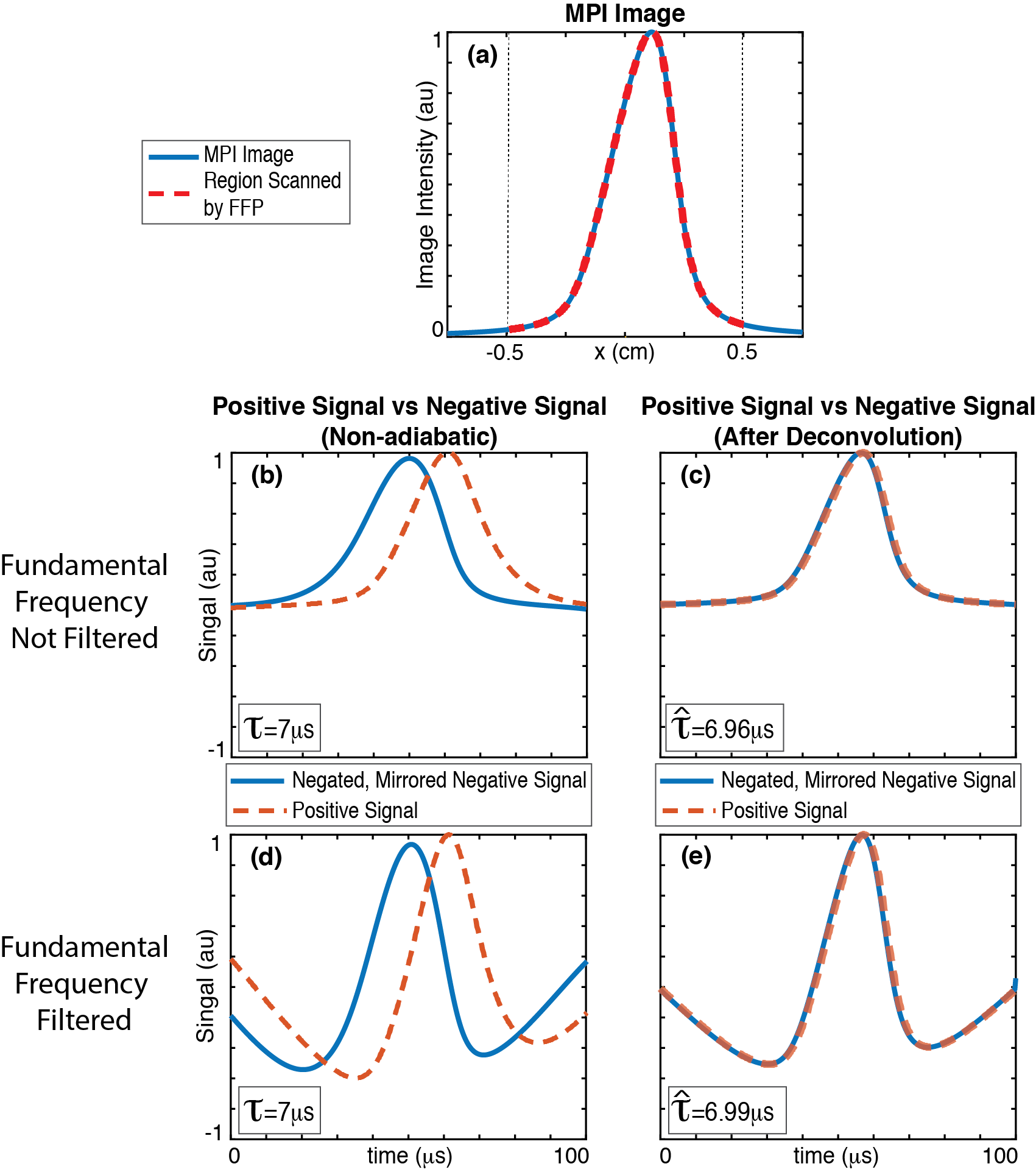}
\caption{The proposed method: direct relaxation time constant estimation.
(a) A simulated 1-D MPI Image (blue) and the covered pFOV extent (red-dashed). (b) The mirror symmetry between positive and negative signals is broken due to relaxation. The proposed method estimated $\hat{\tau} = 6.96~ { \mu}s$., where $\tau = 7~ { \mu}s$ in theory. (c) MPI signal is then deconvolved with the estimated relaxation kernel using $\hat{\tau} = 6.96~ { \mu}s$, which recovers the underlying  mirror symmetry. Here, (b, c) demonstrate the case where the fundamental frequency is kept intact in simulations, and (d, e) demonstrate the case after direct feedthrough filtering that removes the fundamental frequency. In the latter case, the estimation resulted in $\hat{\tau} = 6.99~ { \mu}s$.}
\label{Theoretical_example}
\end{figure}

Figure \ref{Theoretical_example} demonstrates the proposed relaxation time constant estimation method on simulated MPI signal. Here, a triangular nanoparticle density is assumed and a central pFOV is scanned to obtain the MPI signal. Then, relaxation effect is simulated via the exponential model in Eq. \ref{nonadiabatic_signal_equation}, with $\tau = 7~ {\mu}s$. 
Figure \ref{Theoretical_example}b shows the asymmetric blurring due to relaxation, for which the estimated time constant yielded $\hat{\tau} = 6.96~ { \mu}s$. The signal was then deconvolved using the estimated relaxation kernel, recovering the underlying mirror symmetry (see Fig. \ref{Theoretical_example}c). The same procedure was repeated on signal after direct feedthrough filtering \cite{KuanLu2013, Konkle2015}, to ensure that the filtering of the fundamental harmonic (a necessary step in MPI) does not hinder the performance of the proposed technique. The filtered signal is shown in Fig. \ref{Theoretical_example}d with the corresponding deconvolved signal in Fig. \ref{Theoretical_example}e. Here, the estimated time constant was  $\hat{\tau} = 6.99~ { \mu}s$. In the absence of noise in both cases, the proposed method estimated $\tau$ with less than 0.6\% error, which is primarily caused by the digitization in simulating the relaxation effect via convolution.

\section{Materials and Methods}

\subsection{Experimental Relaxation Time Constant Estimation}
Correct timing of $s_{pos}(t)$ and $s_{neg}(t)$ is crucial for the proposed relaxation time estimation method. In simulations, this procedure is trivial since the timing of the signal vs. the FFP trajectory (i.e., the time point, $t_{edge}$, when the positive scan ends and the negative scan starts) is known. In practice, however, system delays introduce a time shift between the signal vs. the FFP trajectory. In such cases, incorrect $t_{edge}$ values introduce extra phase terms in Eq. \ref{tau_estimtaion}, causing incorrect estimation of $\tau$. One method for measuring $t_{edge}$ is performing a calibration scan with an SPIO that does not exhibit relaxation, such that the zero-crossing of the signal acquired during calibration scan can be assigned as $t_{edge}$. Here, we propose a procedure that estimates the correct $t_{edge}$ and $\hat{\tau}$ simultaneously, eliminating the need for a calibration scan.

\begin{figure}[htbp!]
\centering
\includegraphics[width=\linewidth]{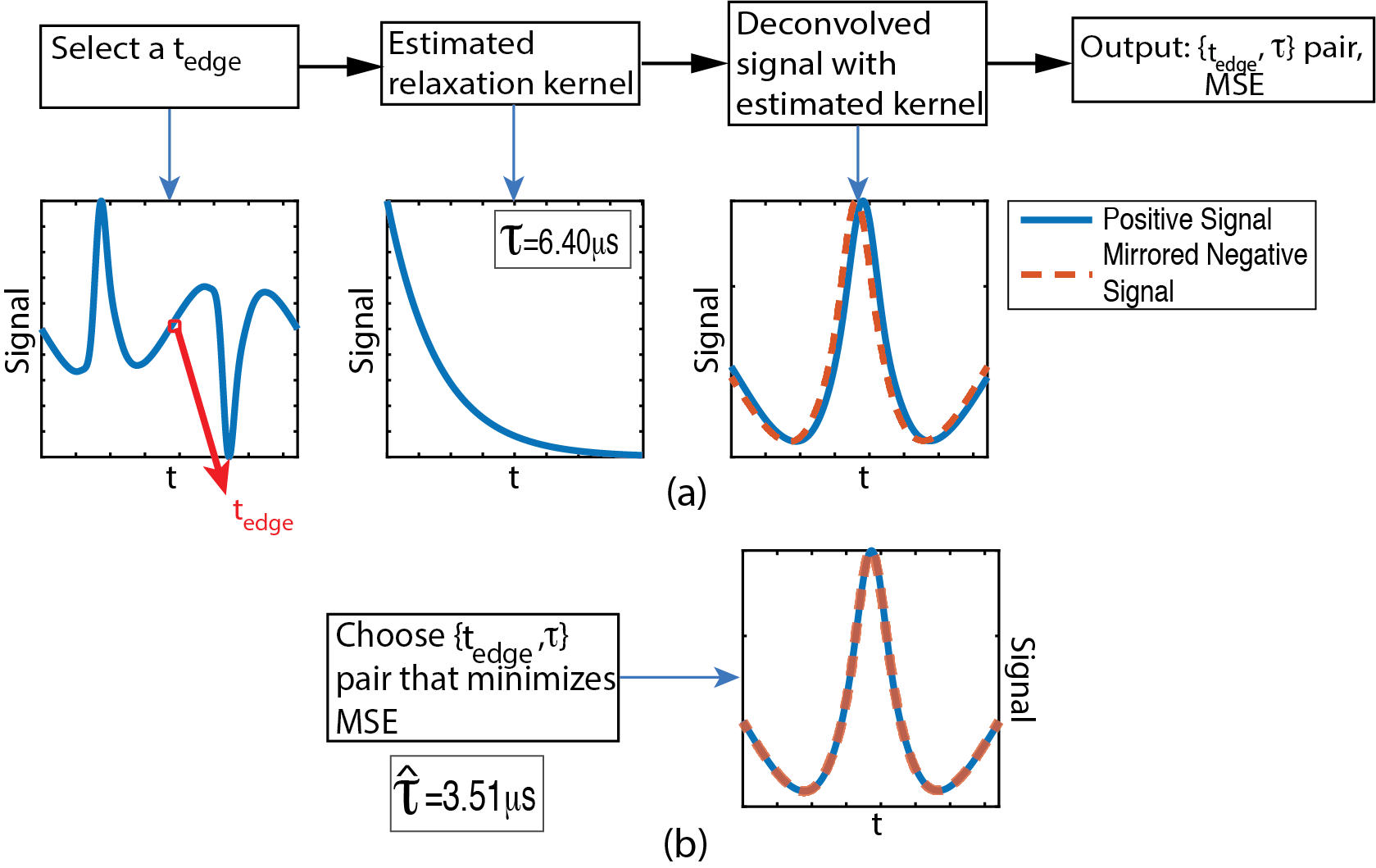}
\caption{Experimental relaxation time constant estimation algorithm. (a) A time point $t_{edge}$ is chosen to mark the time when positive scan finishes and negative scan starts. Then, $\tau$ is estimated for different $t_{edge}$ values chosen from [0, T/2) interval. The MSE between the deconvolved positive and mirrored negative signals are computed for each case. (b) The MSE values are compared, and the $\{t_{edge},\tau\}$ pair that minimizes MSE is chosen as the solution. For this example, the result is $\hat{\tau} = 3.51~ { \mu}s$.
Here, the black arrows show the flow of the algorithm, 
while blue arrows show a visual demonstration of each step.}
\label{Estimation_chart}
\end{figure}

We have argued previously that deconvolution of the MPI signal with the correct relaxation kernel should restore mirror symmetry. Here, we exploit this phenomenon to simultaneously find the correct $t_{edge}$ and $\hat{\tau}$, as outlined in Fig. \ref{Estimation_chart}.
First, we choose a $t_{edge}$ value, directly estimate $\tau$ using Eq. \ref{tau_estimtaion2}, and deconvolve the MPI signal with the estimated relaxation kernel. We then repeat this step for a range of $t_{edge}$ values limited to [0, T/2) region of the MPI signal, where T is the period of the FFP trajectory. Correct ($t_{edge}, \tau$) pair should restore the mirror symmetry, minimizing the mean squared error (MSE) between positive and mirrored negative signals after the deconvolution, i.e.,
\begin{equation}
\label{tau_estimtaion_experimental}
(\hat{t}_{edge}, \hat{\tau})=\argmin_{({t}_{edge}, {\tau})} \int_{0}^{T/2} (\hat{s}_{pos}(t)-(-\hat{s}_{neg}(-t))^2 \ dt
\end{equation}
where $\hat{s}(t)$ denotes the signal after deconvolution with the estimated relaxation kernel. Here, instead of computing MSE over the entire half period, more weights can be assigned to central time points that typically have higher SNR \cite{Utkur2017}.

\subsection{Proposed Algorithm: Calibration-Free Multi-Color MPI}
\label{algorithm}
For the proposed multi-color MPI technique, we directly estimate the relaxation time constant for each pFOV, and map the estimated time constant to the corresponding spatial location. While this mapping sounds simple at first, extensive simulations revealed two special cases where the additional steps are needed for accurate mapping of time constants:

\begin{enumerate}
\item \emph{Flat Nanoparticle Distribution:} If the nanoparticle distribution is flat in a pFOV, most of the signal is contained in the lost 1\textsuperscript{st} harmonic. Hence, direct feedthrough filtering removes almost all signal, making flat regions appear like regions devoid of nanoparticles. In the absence of signal, proposed estimation method produces noise-like results.
\item \emph{Inhomogeneous Mixtures of Different Nanoparticle:} In the regions where two different nanoparticle types mix homogeneously, the estimation yields a weighted average of two relaxation time constants (see Appendix). However, in the case of an inhomogeneous mixture (e.g., a transition region from one nanoparticle type to the other), the adiabatic signal from each nanoparticle not only has a different shape, but is also convolved with a different relaxation kernel. Therefore, the set of equations derived in Eqs. \ref{fourier_pos}-\ref{fourier_neg} become underdetermined, producing noise-like estimations. 
\end{enumerate}

\begin{figure}[htbp!]
\centering
\includegraphics[width=3in]{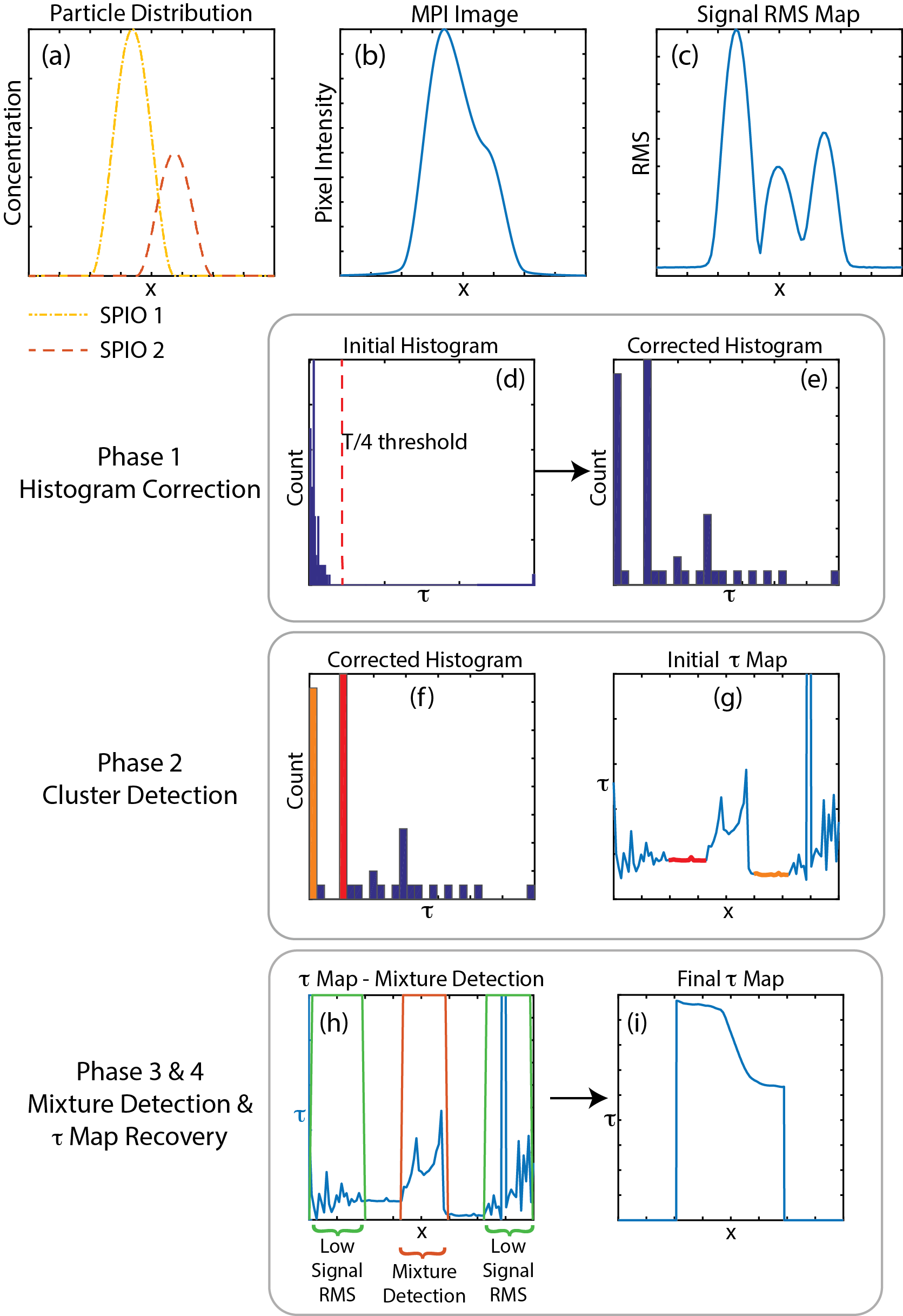}
\caption{The proposed relaxation mapping algorithm. (a) The simulated particle distributions for two different SPIO types, which overlap in a small region in 1-D space. (b) The simulated x-space MPI image. (c) Signal RMS values for each pFOV show that the signal power is reduced in regions where the two SPIOs mix. (d) The initial relaxation time constant estimations are plotted as a histogram, and (e) the histogram is corrected by eliminating low SNR regions and overestimations (i.e., an upper threshold set at T/4). (f) A k-means clustering detects two main clusters (shown in orange and red). (g) These regions are highlighted in the initial $\tau$ map. (h) The inhomogeneous nanoparticle mixture region is detected, and (i) the final $\tau$ is achieved via modeling the transition using a sigmoid curve and setting low pixel intensity regions to zero.}
\label{Mapping_chart}
\end{figure}

Here, we first detect and isolate these special cases, reconstruct the rest of the relaxation map, and finally restore the missing portions of the map. The following steps summarize the proposed multi-color MPI algorithm, outlined in Fig. \ref{Mapping_chart}, that can handle these special cases:\\
\underline{Phase 0 - Estimation of $\tau$ for Each pFOV:} Using the signals from each pFOV, algorithm directly estimates the corresponding relaxation time. An x-space MPI image is also reconstructed (Fig. \ref{Mapping_chart}b), and the signal RMS values for pFOVs are calculated (Fig. \ref{Mapping_chart}c).\\
\underline{Phase 1 - Histogram Correction:} Using the histogram of estimated $\tau$ values, a mild thresholding is applied to remove unlikely estimations 
The histogram correction consists of 2 steps (see Fig. \ref{Mapping_chart}d-e):
\begin{itemize}
\item Upper limit on estimations: Extensive work in nanoparticle relaxometry showed that relaxation time constants are much smaller than the period of the drive field \cite{Croft2015}, typically less than 10\% of the period \cite{Utkur2017}. Here, we ignore estimations that are larger than one-fourth of the period.
\item Signal RMS threshold: Low signal leads to inaccurate $\tau$ estimations. 
Here, estimations from low RMS regions (e.g., less than 10\% of the maximum RMS value) are initially ignored.
\end{itemize}
\underline{Phase 2 - Cluster Detection:} At the end of Phase 1, unlikely estimations are mostly removed. Accordingly, the remaining estimations must have the relaxation times of different SPIO types. Here, a k-means clustering algorithm is run on the remaining estimations \cite{Macqueen1967}. Since the number of nanoparticle types is unknown, $k$ parameter is started from a large value and iteratively reduced until convergence is achieved. Reduction method is as follows (see Fig. \ref{Mapping_chart}f-g):
\begin{itemize}
\item Merger of clusters: Cluster centers that are closer than a $\tau_{res}$ value are merged. Note that $\tau_{res}$ must be sufficiently small to ensure the separation of different clusters.
\item Omission of small clusters: Clusters that contain less than $p\%$ of the overall estimation number are considered irrelevant and ignored. 
\item Convergence: If the cluster properties (number of clusters, cluster centers, etc.) remain the same for two consecutive iterations, convergence is achieved.
\end{itemize}
\underline{Phase 3 - Special Case Detection:} After Phase 2, clusters are detected, which means probable SPIO types and their relaxation time estimations are identified. Outlier $\tau$ values that are not in the vicinity of the cluster centers could indicate either one of the two special cases explained above. These cases have high pixel intensity in the MPI image despite having low signal RMS. Here, we search for the nearest spatial locations that have $\tau$ values belonging in a cluster, on either side of the problem pFOV. If incorrect estimations are caused by \textit{inhomogeneous mixtures}, we expect different clusters on either sides of the problem pFOVs. On the other hand, \textit{flat nanoparticle distributions} have a single cluster type around the problem pFOVs (see Fig. \ref{Mapping_chart}h).\\
\underline{Phase 4 - Recovery of $\tau$ Map:} At the end of Phase 3, the regions of incorrect estimations are determined. These regions are recovered as follows (see Fig. \ref{Mapping_chart}i):
\begin{itemize}
\item For \textit{Flat Nanoparticle Distributions}, the missing regions are recovered as a constant $\tau$ value.
\item For \textit{Inhomogeneous Mixtures}, the average signal RMS of the clusters are used to estimate the relative concentrations of the corresponding nanoparticle types. Next, a transition region from one relaxation time to the other is modeled as a sigmoid function, with steepness determined via relative concentrations.
\end{itemize}

Finally, estimations from low MPI pixel intensity regions are set to zero to remove the background in the $\tau$ map.

\subsection{1D and 3D Simulations}

All simulations were carried out using a custom MPI toolbox developed in MATLAB (Mathworks, Natick, MA). To obtain realistic results, fundamental harmonic was filtered out \cite{KuanLu2013}, relaxation time constant values were selected according to previous studies \cite{Croft2015, Utkur2017}, and smoothly-varying Hamming-window-shaped SPIO distributions were used. The regular MPI images were reconstructed using the x-space reconstruction technique \cite{Goodwill2010, Goodwill2011, KuanLu2013}. 

For 1D simulations, a selection field with 3 $[T/m/\mu_{0}]$ gradient was utilized, with 10 mT-peak drive field at 10 kHz and 80\% overlap between consecutive pFOVs. The nanoparticle diameter was selected as 21 nm, a conservative value given the recent developments in tailored SPIO production \cite{Ferguson2015}. The relaxation times ranged between 1-5 $\mu s$, as recently reported in experimental studies performed around 10 kHz \cite{Croft2015, Utkur2017}. To match experimental conditions, the simulated MPI signal was sampled at 2 MSPS sampling frequency, and additive white Gaussian noise corresponding to a peak signal SNR of 100 was added to the MPI signal. 

For 3D simulations, a selection field with (-7, 3.5, 3.5) $[T/m/\mu_{0}]$ gradient in (x, y, z) directions was utilized, based on the work in \cite{KuanLu2013}. A 15 mT-peak drive field at 10 kHz and a 90\% overlap along z-direction was utilized, with all other parameters kept the same as in 1D simulations. 


\subsection{MPS Experiments}
The initial experiments of proposed relaxation time estimation method were conducted on our in-house Magnetic Particle Spectrometer (MPS, also known as MPI Relaxometer). The drive coil of the relaxometer produces 0.97 mT/A magnetic field with 95\% homogeneity in a 7 cm long region. The receive coil of the relaxometer was designed as a three-section gradiometer pick-up coil \cite{Utkur2015}. 

Experiments were conducted at 550 Hz and 15 mT-peak drive field. The nanoparticle signal was first amplified with a low-noise voltage preamplifier (Stanford Research Systems SR560), then digitized at 2 MSPS via a data acquisition card (National Instruments, NI USB-6363). 16 consecutive acquisitions were averaged to increase the SNR. A background measurement was subtracted from the averaged signal to minimize the effect of direct feedthrough. A frequency domain filter was applied by selecting the higher harmonics of the drive field and setting the rest of the frequency components to zero in Fourier domain. Next, a high-order zero-phase digital low pass filter (LPF) was applied in time domain to restrict the range of harmonics used. The cut-off frequency of the LPF was determined by comparing noise power with signal power at the harmonic frequencies, while avoiding the 250 kHz self-resonance frequency of the receive coil. A more detailed explanation of the complete setup can be found in \cite{Utkur2017}.


\subsection{Imaging Experiments}
\begin{figure}[htbp!]
\centering
\includegraphics[width=\linewidth]{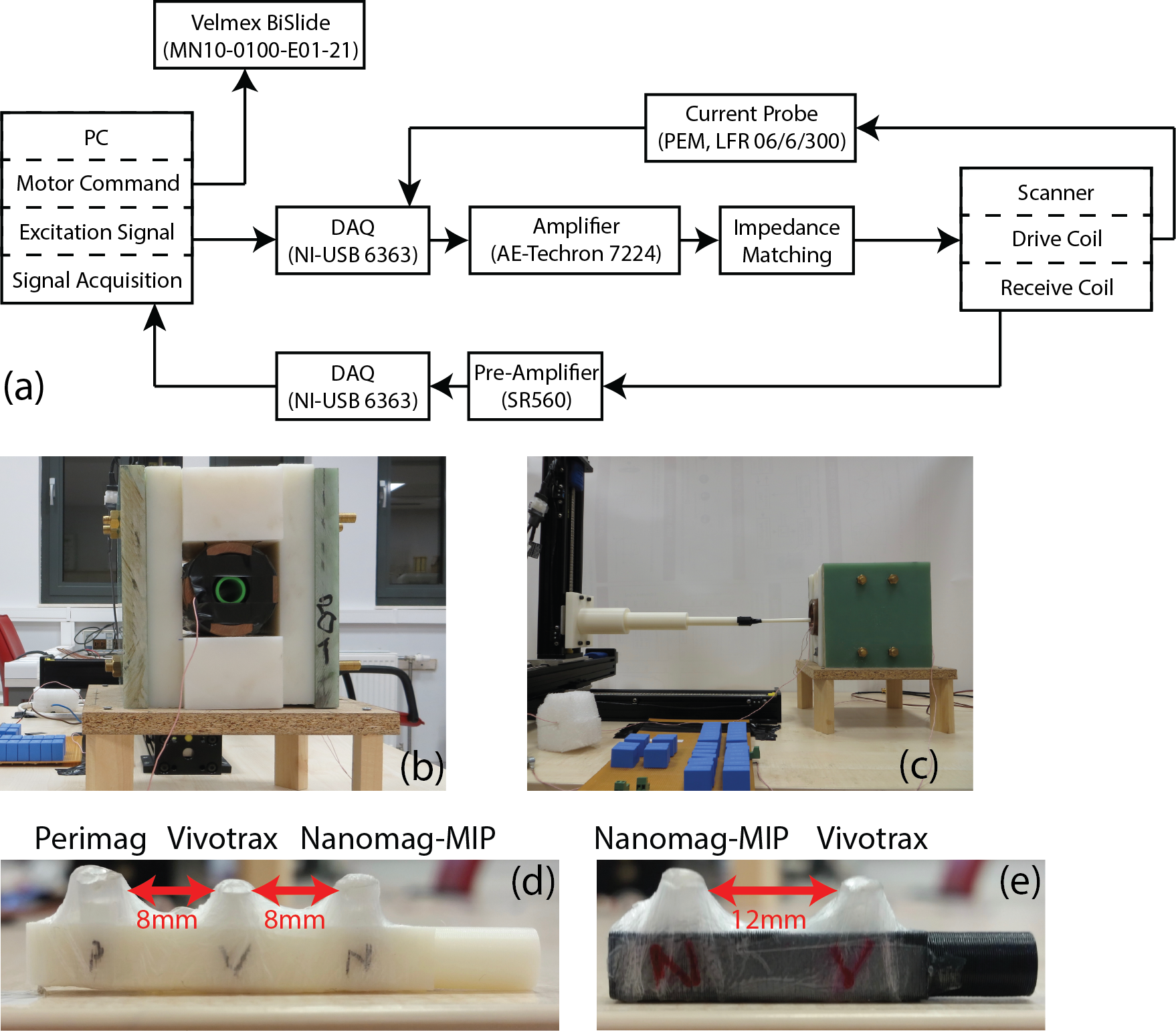}
\caption{An overview of the MPI scanner and the experimental setup. (a) The entire imaging system is controlled through MATLAB, via a custom MPI imaging toolbox. First, robot position is adjusted to place the phantom in the desired location. Second, a sinusoidal drive field at 9.7 kHz is applied through the transmit chain. Simultaneously, the nanoparticle signal is acquired through the receive chain. (b-c) Front and side views our in-house MPI scanner. The phantoms imaged in (d) 1D experiments and (e) 2D experiments.}
\label{Scanner_flow_chart}
\end{figure}

Our in-house MPI Scanner (see Fig. \ref{Scanner_flow_chart}) has two disc-shaped permanent magnets with 2-cm thickness and 7-cm diameter, placed at 8-cm separation in x-direction. The resulting configuration creates (-4.8, 2.4, 2.4) $[T/m/\mu_{0}]$ gradient in (x, y, z) directions, which yields approximately 4-mm resolution in z-direction (i.e., down the imaging bore) for Nanomag-MIP nanoparticles. The drive coil has 3 layers of Litz wire with 80 turns, resulting in 1.5 mT/A magnetic field with 95\% homogeneity in a 4.5 cm-long region down its bore. The drive field and selection field specifications were validated using a Hall Effect Gaussmeter (LakeShore 475 DSP Gaussmeter). The receive coil was designed as a three-section gradiometer, with a single layer of Litz wire with 34 turns in the main section and 17.5 turns in the side sections. The self-resonance of the receive coil was measured at around 280 kHz. This configuration allowed 1x1x10 cm$^3$ FOV. The drive coil and receive coil were placed inside a cylindrical copper shield with 1-cm thickness that can be used for imaging at drive field frequencies as low as 1 kHz.

Figure \ref{Scanner_flow_chart}a-c display the workflow of the complete MPI scanner setup, together with front and side views. Overall imaging system was controlled via a custom toolbox implemented in MATLAB (Mathworks, Natick, MA). The drive coil was impedance matched to AE Techron 7224 power amplifier using a capacitive network at 9.7 kHz. The imaging phantoms were mechanically moved inside the scanner via Motor-Driven Velmex BiSlide (Model: MN10-0100-E01-21) in 3 dimensions. 

The drive field was at 15 mT-peak and 9.7 kHz, which resulted in a 12.5 mm pFOV length in z direction. The amplitude of the drive field was calibrated immediately before each experiment via a Rogowski current probe (LFR 06/6/300, PEM Ltd.). 
Partial FOVs were acquired with 85\% overlaps. For 1D imaging experiments, an 8-cm FOV along z-direction was covered with 16.8 sec active scan time. For 2D imaging experiments, a recti-linear trajectory was used to cover a 0.8x6.8 cm\textsuperscript{2} FOV in x-z plane using 9 lines along z-direction, with 134 sec active scan time. The temperature inside the scanner bore was controlled throughout the experiment to prevent heating of the nanoparticles. 


To boost the SNR, 16 consecutive acquisitions were averaged for each pFOV. 
The background measurements acquired before/after each line were subtracted from the nanoparticle signal to remove direct feedthrough. All remaining signal acquisition/processing steps were the same as in the MPS experiments. The regular MPI images were reconstructed using x-space reconstruction \cite{Goodwill2010, Goodwill2011, KuanLu2013}, followed by the proposed multi-color MPI technique.

\subsection{Phantom Preparation}
Imaging phantoms with different types of SPIOs were prepared, using Vivotrax ferucarbotran nanoparticles (Magnetic Insight Inc., USA) with same chemical composition as Resovist, and Perimag and Nanomag-MIP nanoparticles (Micromod GmbH, Germany). These nanoparticles had original concentration levels of  5.5 mg Fe/mL, 17 mg Fe/mL, and 5 mg Fe/mL, respectively. To approximately equalize the MPI signal levels, Perimag and Nanomag-MIP SPIOs were diluted 10.2 and 3.5 times, respectively. Capillary tubes with 2 mm inner diameter were filled with 10 $\mu$L volume (after dilution) of SPIOs each. These tubes were then placed in custom-designed 3D-printed phantom holders in two different configurations, as shown in Fig. \ref{Scanner_flow_chart}d-e. 


\section{Results}

\subsection{1D and 3D Simulation Results}
\begin{figure}[htbp!]
\centering
\includegraphics[width=2.5in]{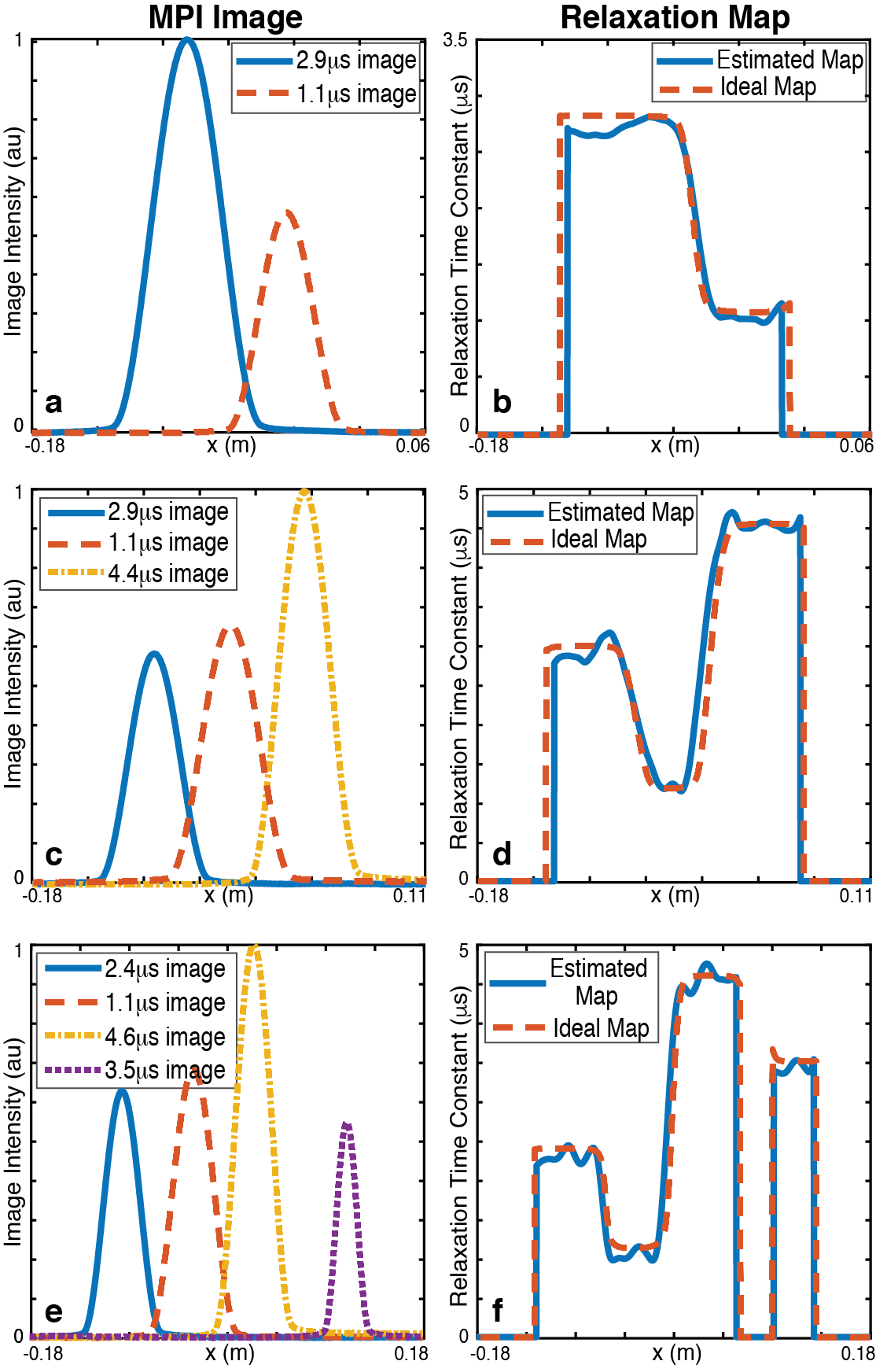}
\caption{Results of 1D simulations for three different cases, for peak signal SNR of 100. Left column (a, c, e) demonstrates different nanoparticle distributions. Right column (b, d, f) demonstrates the corresponding relaxation time constant maps reconstructed with the proposed algorithm.}
\label{1D_results}
\end{figure}

Figure \ref{1D_results} displays the results of the proposed mapping algorithm for the 1D case, where each row corresponds to a different scenario: two, three, and four different types of SPIOs, respectively. The left column shows the regular MPI images for SPIO with relaxation times ranging between 1-5 $\mu s$. The overall MPI image is a summation of these images (not displayed for clarity). Each simulation includes a challenging case of inhomogeneous mixtures of different nanoparticle types. The right column shows the reconstructed relaxation maps (blue solid lines) and ideal relaxation maps (orange dashed lines). While utilizing only two different SPIO types is more likely in medical applications (e.g., catheter tip vs. blood pool in the case of cardiovascular interventions \cite{Rahmer2017}), these results aim to show the full potential of the proposed method.

As seen in each row in Fig. \ref{1D_results}, proposed mapping algorithm reconstructed the relaxation maps accurately, despite the presence of noise. The inhomogeneous mixture regions were detected successfully and the proposed sigmoid-transition model recovered those regions accurately. Furthermore, the proposed algorithm also eliminated background areas via a thresholding and edge detection step. The slight deviations from the ideal relaxation maps were due to noise, as well as the crosstalk of signals from different SPIOs due to proximity. For the results in Fig. \ref{1D_results}b, the mean estimation error is well below 3\%, whereas the mean error reaches 7\% for the results in Fig. \ref{1D_results}f where the SPIO distributions are closer to each other. Here, the 3.5-$\mu$s SPIO was spatially disconnected from the other SPIOs, which is successfully depicted in the reconstructed relaxation map.

\begin{figure}[htbp!]
\centering
\includegraphics[width=3in]{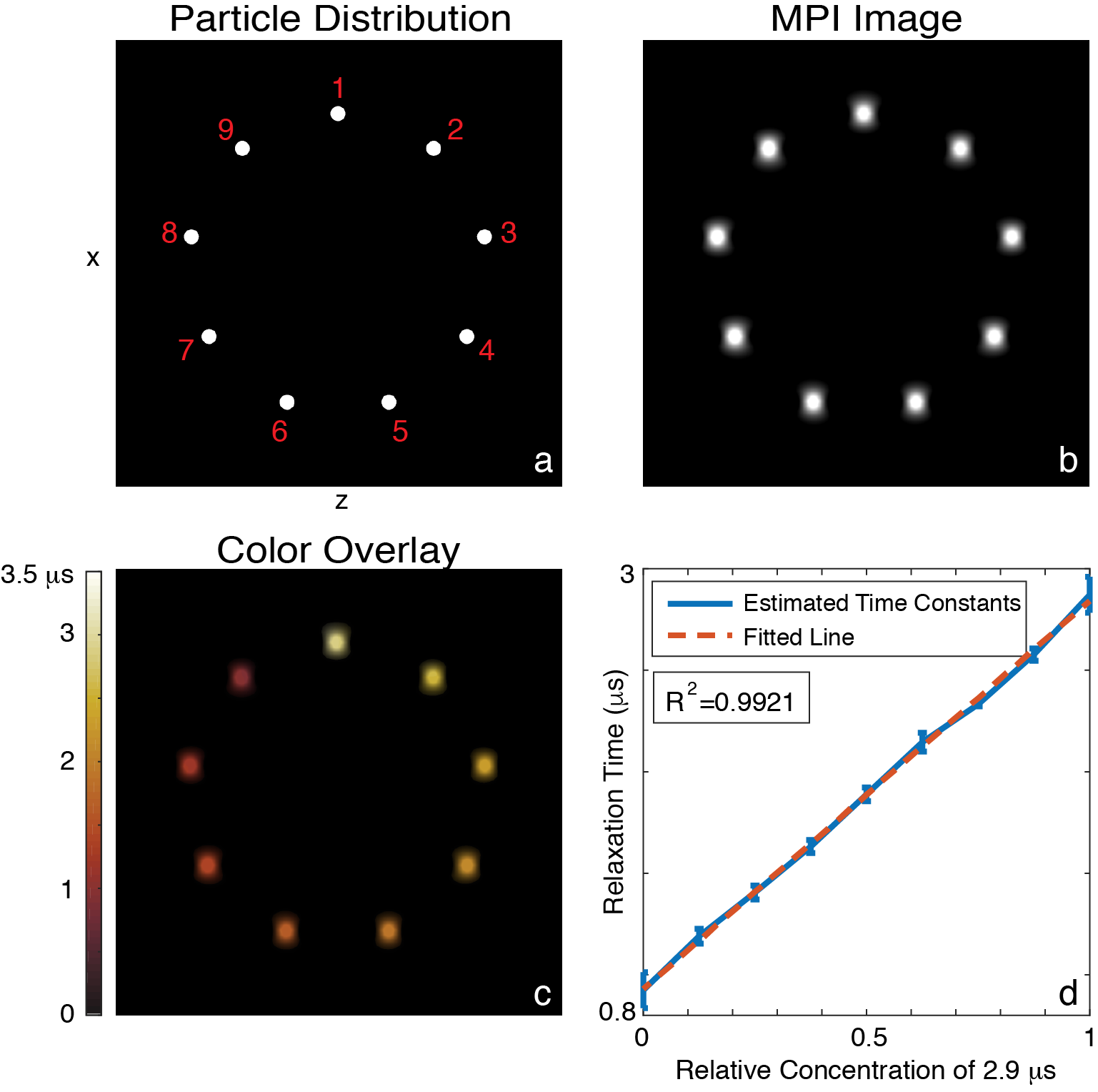}
\caption{Results of 2D simulations, for peak signal SNR of 100. (a) A distribution with homogeneous mixtures of 2.9 $\mu$s and 1.1 $\mu$s SPIOs in different concentrations. Each mixture is labeled with a number from 1 to 9, with concentrations of 2.9 $\mu$s SPIO in each mixture given as (100, 87.5, 75, 62.5, 50, 37.5, 25, 12.5, 0)\%, respectively. 
(b) The x-space MPI image, and (c) the color overlay of the multi-color relaxation map and the MPI image. (d) The estimated time constants vs. relative concentrations of the two SPIOs. Here, the error bars denote the standard deviations for each ROI, and the red dashed curve is the fitted line (with $R^2=0.9921$). FOV size is 9x9 cm$^2$.}
\label{2D_results}
\end{figure}

For the multidimensional case, the proposed technique is extended in a line-by-line basis, where each line in z-direction is reconstructed individually, then combined to form the multidimensional multi-color MPI image. In Fig. \ref{2D_results}a, each SPIO distribution (labeled from 1 to 9) represents a homogeneous mixture of two different SPIOs with 2.9 $\mu$s and 1.1 $\mu$s relaxation time constants at different mixture ratios. Accordingly, the distribution labeled as 1 has 100\% of 2.9-$\mu$s SPIOs, whereas the distribution labeled as 9 has 100\% of 1.1-$\mu$s SPIOs. Figure \ref{2D_results}b-c show the resulting regular MPI image and the color overlay of relaxation map, respectively. In the regular MPI image, the regions corresponding to larger $\tau$ resulted in slightly lower pixel intensities due to relaxation-induced signal loss. However, this difference cannot be used to distinguish the SPIOs, as lower iron concentration could also result in reduced pixel intensity. As seen in Fig. \ref{2D_results}c, the proposed multi-color MPI method is capable of distinguishing a variety of SPIO types (9 in this case), each with a different time constant. Furthermore, the estimated time constants reflect the relative concentrations of the constituent SPIO types with high level of linearity ($R^2=0.9921$), as shown in Fig. \ref{2D_results}d.  

\subsection{MPS Experiment Results}
\begin{figure}[htbp!]
\centering
\includegraphics[width=\linewidth]{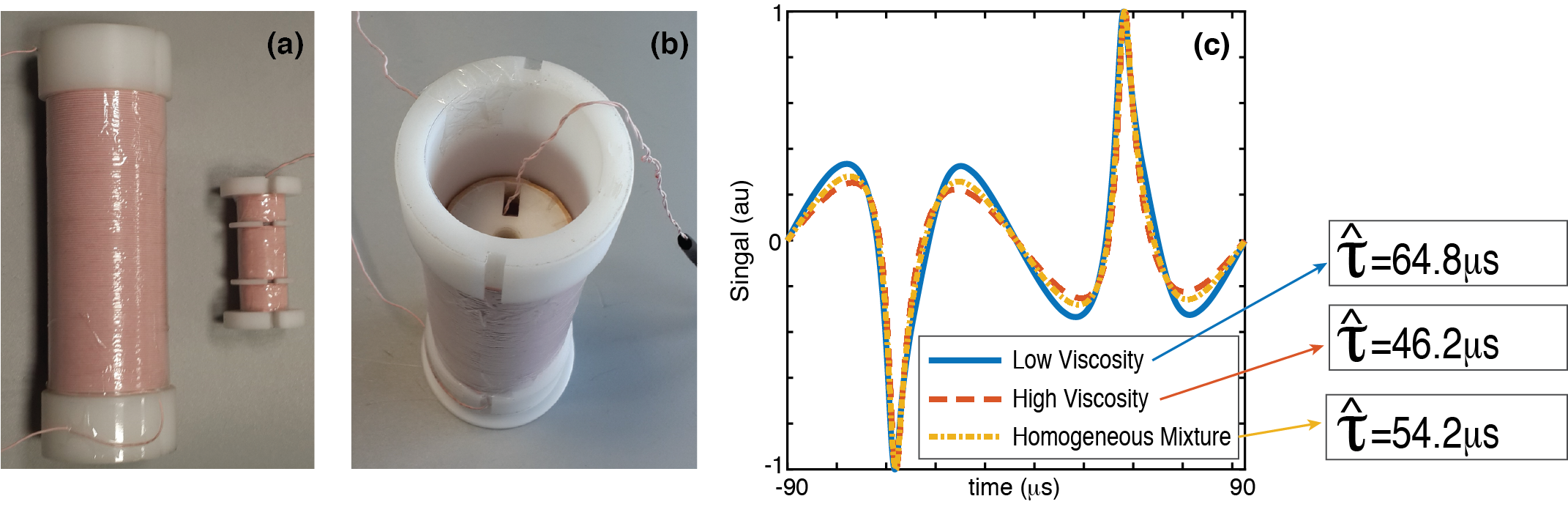}
\caption{Experimental demonstration in MPS setup. (a, b) Pictures of our in-house MPS setup, showing the drive and receive coils placed co-axially.  (c) A  homogeneous  mixture  was  mimicked  by  acquiring signal from two sample tubes simultaneously. The blue and orange curves show the MPI signals of each sample when measured separately, yielding $\hat{\tau} = 64.8~{ \mu}s$ and $\hat{\tau} = 46.2~{\mu}s$. The yellow curve is the MPI signal of the simultaneous acquisition with $\hat{\tau} = 54.2~{\mu}s$, which is approximately equal to the mean of the two time constants.}
\label{relaxometer_results}
\end{figure}

Figure \ref{relaxometer_results}c provides proof-of-concept validation for the simulations given in Fig. \ref{2D_results}. 
In the MPS experiments, nanomag-MIP particles were mixed with water/glycerol at two different ratios, to obtain two different relaxation times due to differences in viscosity levels \cite{Utkur2017}. As given in Fig. \ref{relaxometer_results}c, relaxation time constants for these two cases were estimated as 64.83 $\mu$s and 46.22 $\mu$s. 

Next, a homogeneous mixture was mimicked by acquiring signal from both samples simultaneously. From theory (see Appendix), the resulting relaxation time constant should be the average of the relaxation time constants of the constituents weighted by their concentrations. Here, since the two samples had the same concentration of SPIOs, the expected relaxation time of the homogeneous mixture is 55.52 $\mu$s. The estimated relaxation time of 54.21 $\mu$s closely mathces this expectation.

\subsection{Imaging Experiment Results}
\begin{figure}[htbp!]
\centering
\includegraphics[width=2.5in]{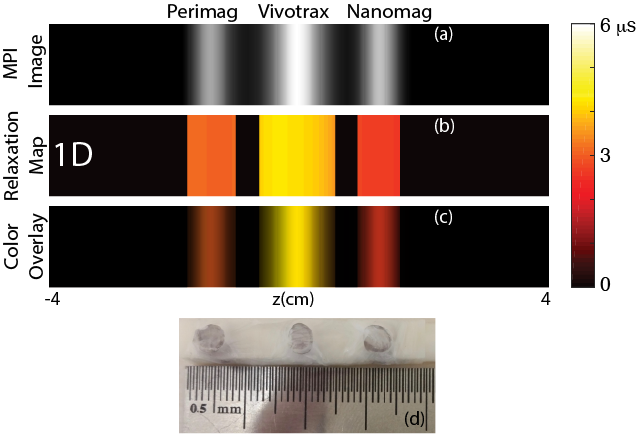}
\caption{Calibration-free multi-color MPI results in 1D. The imaging experiment was conducted at 9.7kHz and 15mT-peak drive field, using three different nanoparticles. (a) 1D reconstructed regular x-space MPI image, and (b) the corresponding multi-color relaxation map. (c) The color overlay of the regular MPI image and the multi-color map shows that the nanoparticles can be clearly distinguished based on their relaxation responses. (d) A photo of the imaged phantom with three tubes containing Perimag, Vivotrax and Nanomag-MIP SPIOs from left to right, separated at 8mm distances.}
\label{1D_scanner_results}
\end{figure}

Figure \ref{1D_scanner_results} displays the 1D imaging results of the proposed technique. In this experiment, a phantom with 3 tubes, each containing a different type of SPIO was imaged. To enable color display, reconstructed 1D regular MPI image and 1D relaxation map were replicated and stacked vertically in a pseudo-2D image format. Figure \ref{1D_scanner_results}a shows the reconstructed regular MPI image. Although the SPIOs were placed in identical tubes, Vivotrax covers a wider region (7.3 mm) in the MPI image, denoting a lower resolution capability. On the other hand, Perimag and Nanomag-MIP SPIOS display comparable widths (5 mm vs. 5.2 mm).

While there are differences between MPI responses of these SPIOs, it is not possible to distinguish the SPIO types based on the regular MPI image in Fig. \ref{1D_scanner_results}a. Figure \ref{1D_scanner_results}b shows the relaxation map reconstructed with the proposed algorithm. Here, different SPIO types can be clearly distinguished after color assignment. On average, Nanomag-MIP, Perimag, and Vivotrax SPIOs yielded 2.7 $\mu$s, 3.5 $\mu$s and 4.5 $\mu$s relaxation time constants, respectively. Finally, Fig. \ref{1D_scanner_results}c is the color overlay image, containing the spatial and quantitative information from both the regular MPI image and the relaxation map.

\begin{figure}[htbp!]
\centering
\includegraphics[width=\linewidth]{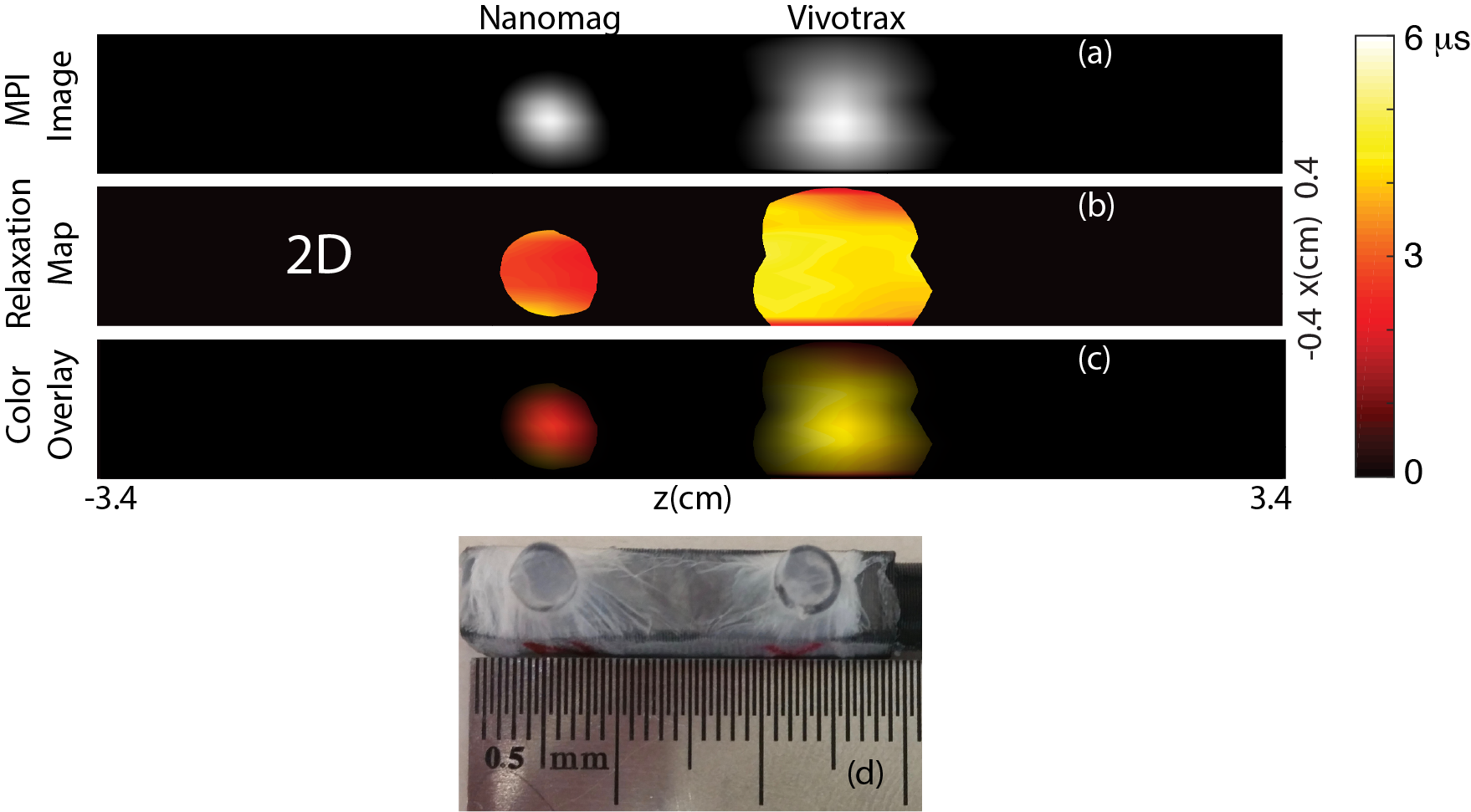}
\caption{Calibration-free multi-color MPI results in 2D. The imaging experiment was conducted at 9.7kHz and 15mT-peak drive field, using two different nanoparticles. (a) The reconstructed regular x-space MPI image and (b) the corresponding multi-color relaxatiom map. (c) The color overlay of the regular MPI image and the multi-color map shows that the two nanoparticles can be clearly distinguished in 2D, solely based on their relaxation responses. (d) A photo of the imaged phantom with two tubes containing Nanomag-MIP and Vivotrax SPIOs from left to right, separated at 12mm distances.}
\label{2D_scanner_results}
\end{figure}

Figure \ref{2D_scanner_results} displays 2D experimental demonstration of the proposed method. Here, Nanomag-MIP and Vivotrax SPIOs were imaged with a 12 mm separation. Again, while the two SPIOs were placed in identical tubes, Vivotrax displays a significantly wider PSF in both x- and z-directions in the regular MPI image in Fig. \ref{2D_scanner_results}a. In both the relaxation map and the color overlay image, Nanomag-MIP and Vivotrax SPIOs can be distinguished clearly. The average time constants were measured as 
2.5 $\mu$s and 4.3 $\mu$s, respectively. Note that these results closely match the relaxation times measured in the 1D experiments for Nanomag-MIP and Vivotrax SPIOs. Here, the colormap (shown in horizontal bar) was chosen to be identical to that in the 1D experiments, to enable a direct comparison of results. Accordingly, 1D and 2D results show a strong agreement, indicating a capability for quantitative imaging using the proposed multi-color MPI technique.

In Fig. \ref{2D_scanner_results}b, at the peripheries of the SPIO distributions along the x-direction, the estimated relaxation times deviate with respect to those in more central regions. While this is an undesired effect of low signal, color overlay image naturally eliminates these low-pixel-intensity regions to provide a clean multi-color MPI image, as seen in Fig. \ref{2D_scanner_results}c.

\section{Discussion}
Multi-color MPI is a very promising extension of MPI due to its potential use in cardiovascular interventions, temperature mapping, and viscosity mapping. There has been significant progress in multi-color MPI, both via SFR and x-space reconstruction techniques. While these methods were shown to work exceptionally well, they either require extensive calibration for every SPIO types used \cite{Rahmer2015} or multiple scans of the imaging FOV at different drive field amplitudes \cite{Hensley2015}. In this work, we have proposed and experimentally demonstrated a calibration-free approach for multi-color MPI, that works successfully with a single scan at a single drive field amplitude. Here, \textit{calibration-free} refers to SPIO identification through their relaxation times, which are estimated \textit{directly} from the MPI signal without any prior information. Note that while relaxation times can be measured in a calibration-free fashion, for certain applications such as viscosity or temperature mapping, one must obtain prior fiducial measurements to create a dictionary of relaxation times corresponding to specific environmental conditions. Alternatively, these measurements could be obtained directly during imaging via placing the fiducials next to the object of interest.

Distinguishable SPIO behavior is a fundamental requirement for any multi-color MPI technique. Rahmer \textit{et al.} have suggested that, SPIOs with similar size distributions or similar hysteresis behavior would be difficult to separate \cite{Rahmer2015}. A similar requirement applies in our case, where SPIOs need to have distinct relaxation time constants. For the experimental results shown in this work, all three nanoparticles involved were multi-core SPIOs, with Perimag and Nanomag-MIP having similar chemical compositions and hydrodynamic diameters. Despite these similarities, the proposed multi-color MPI technique successfully distinguished these nanoparticles. To further improve the performance of this technique, two different approaches can be undertaken: SPIOs that are specifically tailored to have different relaxation behaviors would facilitate relaxation mapping. In addition, imaging parameters can be optimized to maximize the contrast between relaxation times. Our previous work on relaxation-based viscosity mapping showed that relatively low drive field amplitudes and frequencies (e.g., 1 kHz) are more suitable for viscosity mapping, provided that the SNR is sufficiently high \cite{Utkur2017}. 
On the other hand, for distinguishing different SPIO types, it may be more advantageous to operate at higher frequencies and drive field amplitudes. A previous work has compared the relaxation times of multi-core Resovist nanoparticles with single-domain UW33 nanoparticles, which had significantly different relaxation times. Interestingly, the ratio of the relaxation times remained almost constant (around 2-2.5) across a wide range of frequencies and drive field amplitudes \cite{Croft2015}. In such a case, operating at higher fields/frequencies would be desirable for improving the SNR of the MPI signal, although the human safety limits of the drive field should also be taken into account \cite{Saritas2013b, Schmale2013, Saritas2015}.


In this work, we implemented a conventional FFP trajectory, where a piecewise-constant focus field was mimicked with mechanical movement of the robot arm in the x- and z-directions, while a sinusoidal drive field repetitively scanned a chosen pFOV in z-direction. This resulted in 
134 sec active imaging time for the 2D case. The scan times can be significantly shortened by utilizing more efficient FFP trajectories. For example, a high-slew-rate linearly-ramping focus field can be utilized in place of a piecewise-constant focus field. Our preliminary simulation results for such cases yield accurate relaxation time estimations for slew rates as high as 10 T/s (results not shown). With such a trajectory, 
134 sec imaging time would reduce down to approximately 150 ms. Experimental demonstration of rapid imaging with the proposed technique remains a future work.

\section{Conclusion}
In this work, we have proposed a calibration-free multi-color MPI technique and demonstrated the results via both MPS and imaging experiments. The proposed technique relies on distinguishing nanoparticles based on their relaxation response, where the relaxation time is directly estimated from the MPI signal via restoring the underlying mirror symmetry.
A step-by-step algorithm is proposed to generate an accurate relaxation map (or multi-color MPI image) from the time constant estimations for each pFOV. The imaging results show that different nanoparticle responses can be successfully distinguished, without any calibration or prior information regarding their responses. The proposed calibration-free multi-color MPI technique is a promising method for future functional imaging applications of MPI, such as catheter tracking, viscosity mapping, temperature mapping, and stem cell tracking.


\appendix[Relaxation Time Constant of a Homogeneous Mixture]
As stated in Section \ref{algorithm}, when two different SPIOs are mixed homogeneously, $\hat{\tau}$ is the weighted average of the relaxation time constants of the two SPIOs. Here, we assume that the mixture is completely homogeneous and the two SPIOs have identical adiabatic signals,  $s_{half}(t)$ (e.g., bound/unbound SPIOs with identical nanoparticle core sizes). Therefore, Eqs. \ref{nonadiabatic_pos} and \ref{nonadiabatic_neg} can be modified as follows:
\begin{gather}
s_{pos}(t)={c_1}s_{half}(t) * r_1(t)+{c_2}s_{half}(t) * r_2(t) \label{nonadiabatic_pos_mix} \\
s_{neg}(t)={c_1}(-s_{half}(-t) * r_1(t))+{c_2}(-s_{half}(-t) * r_2(t)) \label{nonadiabatic_neg_mix}
\end{gather}
Here, $c_1$ and $c_2$ are the concentrations of the two different SPIOs, and $r_1(t)$ and $r_2(t)$ are the corresponding relaxation kernels with $\tau_1\neq\tau_2$.
Fourier domain representation of Eqs. \ref{nonadiabatic_pos_mix} and \ref{nonadiabatic_neg_mix} can be written as follows:
\begin{gather}
R_i(f)=\frac{1}{1+i2{\pi}f\tau_i}, \:\:   i=1,2 \label{kernel_i} \\
S_{pos}(f)=S_{half}(f).(c_1R_1(f)+c_2R_2(f)) \label{fourier_pos_mix} \\
S_{neg}(f)=-S_{half}^*(f).(c_1R_1(f)+c_2R_2(f)) \label{fourier_neg_mix}
\end{gather}

Here, there are three unknowns, $S_{half}, R_1(f)$ and $R_2(f)$, and two equations, Eq. \ref{fourier_pos_mix} and \ref{fourier_neg_mix}. Hence, the equation set above is underdetermined. Nevertheless, if the original estimation method given in Eq. \ref{tau_estimtaion} is applied, the following expression can be obtained using Eqs. \ref{fourier_pos_mix}-\ref{fourier_neg_mix}.:
\begin{align}
\label{estimation_mix}
\hat{\tau}(f)&=\frac{S_{pos}^*(f)+S_{neg}(f)}{i2{\pi}f(S_{pos}^*(f)-S_{neg}(f))} \nonumber \\
&=\frac{2c_1\tau_1(1+4\pi^2f^2\tau_2^2)+2c_2\tau_2(1+4\pi^2f^2\tau_1^2)}{2c_1(1+4\pi^2f^2\tau_2^2)+2c_2(1+4\pi^2f^2\tau_1^2)} 
\end{align}

Equation \ref{estimation_mix} is not strictly linear with respect to concentrations of the SPIOs in the mixture. However, for small frequencies (e.g., for $f < 4f_0$ where $f_0$ is the drive field frequency), the terms $(1+4\pi^2f^2\tau_i^2)$ can be approximated as 1. In that case, Eq. \ref{estimation_mix} simplifies to the following form:
\begin{equation}
\label{estimation_mix_simple}
\hat{\tau}(f)\approx\frac{c_1\tau_1+c_2\tau_2}{c_1+c_2}
\end{equation}
where $\hat{\tau}(f)$ is equal to the weighted average of $\tau_1$ and $\tau_2$. 

Note that, if the adiabatic responses of the two SPIOs are different (e.g., SPIOs with different core sizes), the estimated $\hat{\tau}(f)$ will be an average of $\tau_1$ and $\tau_2$, weighted by the overall signal levels from the two SPIOs. 


\section*{Acknowledgment}
A preliminary version of this work was presented at the $6^{th}$ International Workshop on Magnetic Particle Imaging (IWMPI 2016). This work was supported by the Scientific and Technological Research Council of Turkey (TUBITAK 114E167), by the European Commission through FP7 Marie Curie Career Integration Grant (PCIG13-GA-2013-618834), by TUBA-GEBIP 2015 program of the Turkish Academy of Sciences, and by BAGEP 2016 Award of the Science Academy.

\bibliographystyle{IEEEtran}
\bibliography{colorMPIbib}
\end{document}